\documentclass[10pt,a4paper]{article} 
\usepackage{jcappub}
\usepackage{times}
\usepackage{graphicx}
\usepackage{comment}
\pdfoutput=0        

\usepackage{color}

\newcommand{\mpc}{{\rm Mpc}}
\newcommand{\gpc}{{\rm Gpc}}
\newcommand{\hmpc}{{h^{-1}\mpc}}
\newcommand{\rms}{\sigma_8}
\newcommand{\mpci}{{\mpc^{-1}}}
\newcommand{\up}[1]{{\rm #1}}
\newcommand{\msun}{M_{\odot}}
\newcommand{\bdv}[1]{{\bf #1}}
\newcommand{\beeq}{\begin{equation}}
\newcommand{\kk}{\bdv{k}}
\newcommand{\TT}{T}
\newcommand{\PC}{\varphi_p}
\newcommand{\eneq}{\end{equation}}
\newcommand{\SO}{\sigma_{rv}}
\newcommand{\PS}{\psi}
\newcommand{\UR}{u_r}
\newcommand{\xx}{\bdv{x}}
\newcommand{\rr}{\bdv{r}}
\newcommand{\PST}{\psi_\perp}
\newcommand{\PSP}{\psi_\parallel}
\newcommand{\PV}{P_u}
\newcommand{\PPH}{P_\varphi}
\newcommand{\PPM}{P_m}
\newcommand{\hmpci}{{h\mpc^{-1}}}
\newcommand{\ccV}{b_r}
\newcommand{\Dh}{\delta_g}
\newcommand{\ccL}{b_1}
\newcommand{\sur}{\sigma_{ru}}
\newcommand{\bear}{\begin{eqnarray}}
\newcommand{\qq}{\bdv{q}}
\newcommand{\enar}{\end{eqnarray}}
\newcommand{\BB}{B_g}
\newcommand{\PX}{P_\times}
\newcommand{\PP}{P_m}
\newcommand{\GG}{G_u}
\newcommand{\FI}{{(1)}}
\newcommand{\SE}{{(2)}}
\newcommand{\TI}{{(3)}}
\newcommand{\FF}{F_2}
\newcommand{\ccN}{b_2}
\newcommand{\ccH}{b_3}
\newcommand{\Ph}{P_g}
\newcommand{\PPN}{P_\up{NL}}
\newcommand{\PPT}{P_t}
\newcommand{\hgpc}{{h^{-1}\gpc}}
\newcommand{\pp}{\bdv{p}}
\newcommand{\CC}{\bdv{C}}
\newcommand{\mm}{{\mbox{\boldmath$\mu$}}}
\newcommand{\dd}{\bdv{D}}
\newcommand{\dmm}{{\mbox{\boldmath$\delta\mu$}}}
\newcommand{\ff}{\bdv{F}}

\title{Supersonic relative velocity effect on the baryonic acoustic
oscillation measurements}

\author[a,b]{Jaiyul Yoo}
\emailAdd{jyoo@physik.uzh.ch}
\author[c]{Neal Dalal}
\author[a,b,d,e]{Uro{\v s} Seljak}

\affiliation[a]{Institute for Theoretical Physics, University of Z\"urich,
CH-8057 Z\"urich, Switzerland}
\affiliation[b]{Lawrence Berkeley National Laboratory, University of 
California, Berkeley, CA 94720, USA}
\affiliation[c]{Canadian Institute for Theoretical Astrophysics, University
of Toronto, 60 St. George Street, Ontario, M5S 3H8, Canada}
\affiliation[d]{Physics Department and Astronomy Department,
University of California, Berkeley, CA 94720, USA}
\affiliation[e]{Institute for the Early Universe, Ewha Womans University, 
120-750 Seoul, South Korea}

\abstract{We investigate the effect of
supersonic relative velocities between baryons and dark matter, 
recently shown to arise generically at high redshift, on 
baryonic acoustic oscillation (BAO) measurements at low redshift. 
The amplitude of the relative velocity effect at low redshift
is model-dependent, but
can be parameterized by using an unknown bias. We find that if unaccounted,
the relative velocity effect can shift the BAO peak position 
and bias estimates of the dark energy equation-of-state due to its
non-smooth, out-of-phase oscillation structure around the BAO scale.
Fortunately, the relative velocity effect can be easily modeled in
constraining cosmological parameters without substantially inflating
the error budget.  We also 
demonstrate that the presence of the relative velocity effect gives rise
to a unique signature in the galaxy bispectrum, which can be 
utilized to isolate this effect.
Future dark energy surveys can accurately measure the relative velocity 
effect and subtract it from the power spectrum analysis to constrain 
dark energy models with high precision.}

\arxivnumber{}

\renewcommand{\beforetochook}{\clearpage}  
\toccontinuoustrue    

\begin{document}
\maketitle
\flushbottom

\section{Introduction}
\label{sec:intro}
Understanding the nature of dark energy is one of the key questions in
contemporary science, and various techniques have been developed
over the past decade
to advance this goal. Measurements of the baryonic acoustic
oscillation (BAO) feature in galaxy surveys provide an
especially promising way to constrain the expansion history of the
universe and the behavior of dark energy \cite{DETF06}.  
The physical scale of the
acoustic oscillation is determined with high accuracy by cosmic microwave
background (CMB) measurements (e.g., \cite{KOSMET11,LADUET11}), 
making BAO measurements at
low redshift a robust standard ruler.  Since current and future dark energy
surveys aim to measure the BAO peak position at sub-percent level
precision, significant efforts have been devoted to modeling nonlinear
evolution and its effect on the BAO peak position
\cite{SEEI05,JEKO06,EISEWH07,CRSC08,SMSCSH08,TANISA10,DECRET10,ORWE11}.
At low redshift, where nonlinear effects are substantial on BAO scales,
the BAO peak contrast is degraded and the peak position may be
subtly shifted compared to the linear theory prediction. However,
these nonlinear effects tend to produce smooth, broad-band power and
may readily be removed by template fitting or explicit modeling of
these effects 
\cite{SESIET08,SABAAN08,YOMI10}. 

Recently, \citet{TSHI10} discussed a new nonlinear effect in the growth of
small scale structure at very early times. Prior to the recombination epoch,
baryons are tightly coupled to the photons, while dark matter is decoupled
from the baryon-photon fluid and its fluctuation grows logarithmically
with small velocity.
After cosmic recombination, when the tight coupling of the baryon-photon
fluid is broken and the gas sound speed plummets, the relative
velocity between the baryon and the dark matter fluids becomes
supersonic. These random relative velocities 
effectively increase the Jeans mass for the formation of the
earliest baryonic structure, thereby suppressing their abundance 
\cite{TSHI10,DAPESE10,TSBAHI10}. High-resolution numerical simulations
\cite{GRWHET11,MAKOCI11} have confirmed that 
supersonic relative velocities raise the minimum halo mass that can form 
the first stars, and delay star formation in time, by an amount that depends
on the magnitude of the relative velocity (see also \cite{STBRLO10}).
This modulation of baryonic structures at high redshift imprints
signatures of the relative velocity effect in the power spectrum of
objects such as minihalos at high redshift \cite{TSHI10,DAPESE10}, 
and hence the power spectrum of any observable which traces these
objects can exhibit {\it significant} departures from simple linear biasing
of the matter power spectrum on BAO ($\sim 100~\hmpc$) scales.

Even at low redshift $z\lesssim5$, when 
massive dark matter halos form, the relative velocity effect can 
be indirectly important, contrary to the naive expectation
that this effect becomes negligible within halos whose velocity
dispersion greatly exceeds the relative velocity.  
One possible scenario \cite{DAPESE10} is that patchy reionization, driven
by early minihalos, inherits the imprints of the relative velocity
effect in the spatial distribution of early minihalos, 
resulting in a spatial variation of the subsequent star 
formation history modulated by the large-scale power of the relative
velocity effect. Consequently, massive galaxies at low
redshift may retain the memory of the large-scale relative velocity 
effect seen in the minihalos at high redshift.  Even tiny $\sim 1\%$
level effects on the colors or luminosities of massive galaxies can 
have significant implications,
since these effects will be correlated across $\sim 100\hmpc$ 
length scales.  Any residual relative velocity effect
in low-redshift massive galaxies may have significant impact on the precision
measurements of the BAO peak position. Since the power spectrum of
relative velocities
has a prominent oscillation structure that is out-of-phase with the matter
power spectrum, it cannot be removed by a blind broad-band fitting of the
galaxy power spectrum, and therefore it can potentially bias the determination
of the BAO peak position. 

The primary purpose of the present work
 is to investigate the impact of any relative
velocity effect that persists in low-redshift massive galaxies, on our ability
to probe the underlying cosmology using galaxy power spectrum measurements,
and to find possible ways to isolate and remove this effect
in the analysis.
The organization of this paper is as follows. In section~\ref{sec:formalism}
we briefly describe the supersonic relative velocity effect and its impact
on the earliest baryonic structure. In section~\ref{sec:bispectrum} we
model the observed galaxy as a tracer of the underlying matter and the
relative velocity distributions and compute the galaxy bispectrum to isolate
the relative velocity contribution. We then investigate the impact of the
relative velocity effect on the BAO measurements in section~\ref{sec:power}.
We first compute the full power spectrum including the relative velocity
effect in section~\ref{ssec:pow} and 
quantify the shift in the BAO peak position
due to the relative velocity effect in section~\ref{ssec:bao}.
In section~\ref{sec:for} we perform a Fisher matrix analysis to investigate
the impact of the relative velocity effect on the cosmological parameter
estimation and how well the relative velocity effect can be measured 
by using the bispectrum. Finally, we conclude in section~\ref{sec:discussion}
with a discussion of further implication.
Technical details of our derivations are presented in Appendices~\ref{ap:com}
and~\ref{ap:bias}.

Here we present our calculations assuming a flat $\Lambda$CDM universe with 
the matter density $\omega_m=\Omega_mh^2=0.134$ ($\Omega_m=0.271$), 
the baryon density $\omega_b=\Omega_bh^2=0.0222$ ($\Omega_b=0.045$), 
the spectral index $n_s=0.966$ 
and its running $\alpha_s=0$ of the
primordial curvature power spectrum $\Delta_\varphi^2=2.42\times10^{-9}$ 
($\rms=0.81$) at $k_0=0.002\mpci$ (consistent with the Wilkinson 
Microwave Anisotropy Probe measurements \cite{KOSMET11,LADUET11}
and the Sloan power spectrum measurements \cite{TESTET06,REPEET09}).

\section{Supersonic relative velocity effect}
\label{sec:formalism}
Before the recombination epoch, baryons and photons are tightly coupled
with sound speed $c_s\simeq1/\sqrt{3}$, 
while dark matter is decoupled and cold. At the
release of baryons following cosmic recombination, the gas sound speed
drops precipitously ($c_s\sim10^{-5}$),
rendering the baryon fluids supersonic ($\mathcal{M}\simeq 2-5$).
The different velocities
of the dark matter and the baryon fluids
result in suppression of the matter power
spectrum at the Jeans scale ($k_\up{J}\sim200~\mpci$) and 
suppression of the growth of
structure at the relative velocity scale 
($k_r=k_\up{J}/\mathcal{M}$, 
$M_\up{h}\simeq10^6\msun\sim k_r^{-3}$) \cite{TSHI10,TSBAHI10}. Furthermore,
the relative velocity effect modulated by large-scale acoustic oscillations
imprints its signature in the collapsed baryon fraction at early time,
and this effect can be used to probe the nature of minihalos before 
reionization \cite{DAPESE10}.

The velocities $\bdv{v}_{b,m}$ of the baryon and dark matter distributions
are related to their density fluctuations as
\beeq
\bdv{v}_{b,m}(\kk,z)=i\frac{\kk}{k^2}~a~\dot\delta_{b,m}(\kk,z)
=i\frac{\kk}{k^2}~a~\dot\TT_{b,m}(k,z)~\PC(\kk) ~,
\label{eq:velo}
\eneq
where $\PC(\kk)$ is the primordial curvature perturbation in the 
total comoving gauge. The evolved
density fluctuations are computed using the Einstein-Boltzmann code 
{\footnotesize CMBFAST} \cite{SEZA96} and are expressed in terms of their
transfer functions $\delta_{b,m}(\kk,z)=\TT_{b,m}(k,z)~\PC(\kk)$. 
Instead of taking
the derivatives of the transfer functions, we simply use the velocity transfer
function $\TT_v(k,z)$ from the {\footnotesize CMBFAST} 
code (it is related to the
density transfer function as $\TT_{vb,vm}=-a\dot\TT_{b,m}/k$).
The relative velocity of the baryon and the dark matter fluids
is then $\bdv{v}_r=\bdv{v}_b-\bdv{v}_m$, and we define the dimensionless
relative velocity as $\bdv{u}_r=\bdv{v}_r/\SO$ (and their transfer 
function~$\TT_{ru}$ as well) using the one-dimensional rms 
relative velocity fluctuation $\SO$.
Their statistical properties are completely determined by the two-point
correlation function
\beeq
\PS_{ij}(r)=\langle\UR^i(\xx)\UR^j(\xx+\rr)\rangle
=\PST(r)~\delta_{ij}+\left[\PSP(r)-\PST(r)\right]\delta_{iz}~\delta_{jz}
=\int{d^3\kk\over(2\pi)^3}~e^{i\kk\cdot\rr}~\PV^{ij}(\kk)~,
\eneq
where the indices~$i,j$ represent the spatial component of the relative
velocity vector, the $z$-direction is set parallel to the separation 
vector~$\rr$ ($\PST=\psi_{xx}=\psi_{yy}$, $\PSP=\psi_{zz}$), and the relative
velocity power spectrum is defined as
\beeq
\langle\UR^i(\kk_a)\UR^j(\kk_b)\rangle=(2\pi)^3\delta^D(\kk_a+\kk_b)
~\PV^{ij}(\kk_a)~
\eneq
with
\beeq
\PV^{ij}(\kk)={k^ik^j\over k^4}\left[{a~\dot\TT_r(k)\over\SO}\right]^2
\PPH(k)={k^ik^j\over k^2}~\TT_{ru}^2(k)~\PPH(k)={k^ik^j\over k^2}~
{\TT_{ru}^2(k)\over\TT_m^2(k)}~\PPM(k)~.
\eneq
For notational simplicity, we suppressed the time dependence of the transfer
functions and their power spectra.

\begin{figure}
\centerline{\psfig{angle=-90, file=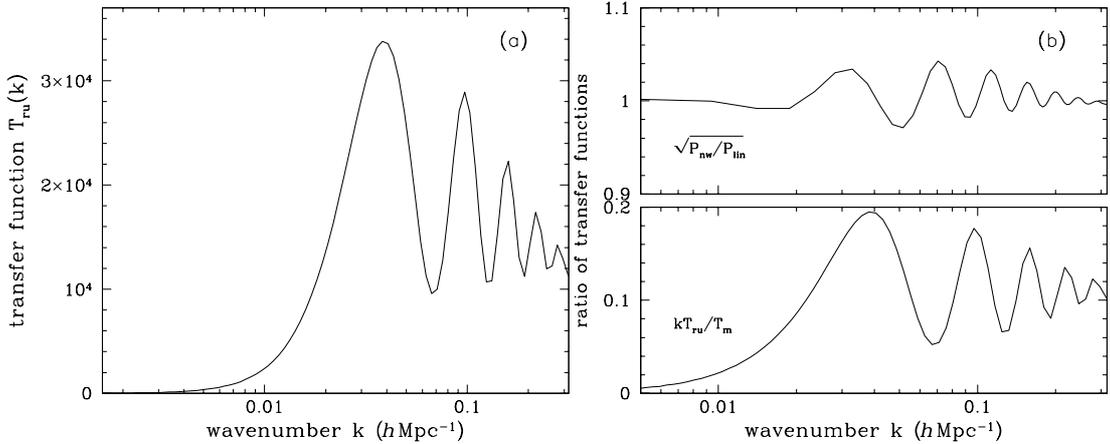, width=\textwidth}}
\caption{
Transfer functions and correlation functions.
($a$)~Relative velocity transfer function $\TT_{ru}(k)$ at $z=0$.
($b$)~Ratio of the relative velocity to the dark matter transfer functions
$k\TT_{ru}/\TT_m$ (bottom panel).
For comparison, the upper panel shows the
ratio of the no-wiggle power spectrum to the linear matter power
spectrum. The transfer functions
are defined with respect to the primordial curvature perturbation,
and the no-wiggle power spectrum is computed by using the \citet{EIHU98}
fit to the smooth power spectrum shape.}
\label{fig:f1}
\end{figure}

Figure~\ref{fig:f1}a shows the transfer function $\TT_{ru}(k)$ of the
relative velocity effect at $z=0$, where its amplitude indicates the linear
growth of power given the primordial curvature power spectrum
$\Delta_\varphi^2(k)=A(k/k_0)^{n_s-1}$ with $A=2.4\times10^{-9}$ and
$k_0=0.002~\mpci$.
The relative velocity is predominantly sourced by structures on size
scales of order the acoustic scale; larger-scale structures
make little contribution, and the transfer function $\TT_{ru}(k)$ declines
rapidly at $k\lesssim0.03\hmpci$ since baryons and dark matter behave
similarly on scales beyond the sound horizon at decoupling.
The ratio of the relative velocity to the dark matter transfer functions
$\TT_{ru}(k)/\TT_m(k)$ is shown in Fig.~\ref{fig:f1}b, where the scaling 
factor~$1/k$ reflecting energy-momentum conservation
is removed. The prominent oscillation
structure displayed in Fig.~\ref{fig:f1}a is still visible 
in the bottom panel of Fig.~\ref{fig:f1}b, when compared
to the matter transfer function. To facilitate the
comparison of the acoustic oscillation amplitude and its phase
in the relative velocity
and the dark matter transfer functions, we compute
the ratio of the no-wiggle power spectrum to the linear matter power
spectrum and plot its square root in the upper panel of 
Fig.~\ref{fig:f1}b,
where the no-wiggle power spectrum is a fit to
the smooth shape of the linear matter power spectrum \cite{EIHU98}.
The oscillation structure in the relative velocity and the matter distributions
is out-of-phase, and its amplitude ($\sim$5\%) is smaller 
in the matter distribution
as the acoustic oscillation was only present in the baryon distribution,
not in the dark matter distribution at early time.
By contrast, the relative velocity arises entirely from the acoustic 
oscillations, and hence its oscillation amplitude is fractionally
order one.

However, the relative velocity of the baryon and the dark matter distributions
is {\it not} 
directly observable. \citet{DAPESE10} argue that the collapsed baryon
fraction at early time is affected by the relative velocity effect as it
changes the effective sound speed and increases the 
equivalent Jeans mass for baryonic
gas to collapse in dark matter halos. The two-point correlation function of the
collapsed baryon fraction can be computed \cite{DAPESE10} as
\beeq
\xi_f(r)=\ccV^2\left[\PST^2(r)+{1\over2}\PSP^2(r)\right]\propto
\left\langle\UR^2(\xx)\UR^2(\xx+\rr)\right\rangle~,
\eneq
implying that the collapsed baryon fraction is a biased tracer of the
relative velocity $\UR^2(\xx)$.
We quantify the effect of the relative velocity on the
BAO peak shift in section~\ref{sec:power}.

\section{Separating relative velocity contribution: Bispectrum}
\label{sec:bispectrum}
At high redshift, $z>10$,
the supersonic relative velocity allows baryons to advect
out of small dark matter halos, effectively increasing the Jeans mass
of the gas in a velocity-dependent manner, thereby modulating the
large-scale clustering of collapsed baryonic objects \cite{DAPESE10}. 
Additionally, the abundance of dark matter halos is also modulated by
a similar effect \cite{TSHI10}.  Quite generally, we can write 
the large-scale fluctuation of the collapsed objects 
as \cite{DAPESE10}
\beeq
\label{eq:basic}
\Dh(\xx)=\ccL~\delta_m(\xx)+\ccV\left[\UR^2(\xx)-\sur^2\right]~,
\eneq
where $\ccL$ and $\ccV$ are the bias parameters of the collapsed halo
with respect to the matter 
density and the relative velocity, respectively. Hereafter,
we collectively
call the collapsed baryon and dark matter system a galaxy, though a 
galaxy is often used to refer to a baryon only system and
at high redshift the collapsed baryon and dark matter systems may appear
different from typical ``galaxies'' at low redshift.
The galaxy fluctuation is
not only a tracer of the underlying matter distribution, but also a tracer
of the relative velocity. The magnitude of the relative velocity bias~$\ccV$
is computed by \citet{DAPESE10} at $z>10$, where the collapse of the 
baryonic structure can be relatively simply modeled. 
As discussed above, the relative velocity between baryons and dark matter can
indirectly modulate the properties of galaxies at low redshift.
The large-scale clustering properties of observed galaxies can, to
lowest order, therefore be written in the form of
eq.~(\ref{eq:basic}).  In this case the velocity bias $\ccV$ is treated as a
free parameter, reflecting our great 
uncertainty in the physical
processes that determine galaxy properties at low redshift.

Since the relative velocity $\UR$
enters quadratically in eq.~(\ref{eq:basic}),
it is readily apparent that the relative velocity effect will lead to a
nonvanishing bispectrum of the galaxy fluctuation, providing a direct way
to isolate its contribution from the matter density distribution.
Noting that the relative velocity has no directional dependence, the ensemble
average of the galaxy fluctuations can be computed as
\bear
\label{eq:three}
&&\hspace{-25pt}
\left\langle\Dh(\kk_a)\Dh(\kk_b)\Dh(\kk_c)\right\rangle 
=\ccL^2~\ccV\int{d^3\qq\over(2\pi)^3}\sum_{i=x,y,z}
\left\langle\Dh(\kk_a)\Dh(\kk_b)\UR^i(\qq)\UR^i(\kk_c-\qq)\right\rangle  
+\up{cyclic~permutations}~  \\
&&\hspace{10pt}+\ccV^3\int{d^3\qq_1\over(2\pi)^3}\int{d^3\qq_2\over(2\pi)^3}
\int{d^3\qq_3\over(2\pi)^3}\sum_{i,j,k}
\left\langle\UR^i(\qq_1)\UR^i(\kk_a-\qq_1)
\UR^j(\qq_2)\UR^j(\kk_b-\qq_2)
\UR^k(\qq_3)\UR^k(\kk_c-\qq_3)\right\rangle  ~. \nonumber
\enar
Counting only the connected part ($\kk_a+\kk_b+\kk_c=0$)
in eq.~(\ref{eq:three}), we can derive the galaxy bispectrum (see 
Appendix~\ref{ap:com})
\bear
\label{eq:bvl}
&&\BB(\kk_a,\kk_b,\kk_c)
=2~\ccL^2~\ccV\!\!\sum_{i=x,y,z}\!\!\left[
\PX^i(\kk_a)\PX^i(\kk_b)+\up{cycl.} \right]
+8~\ccV^3\int{d^3\qq\over(2\pi)^3}\sum_{i,j,k}
\PV^{ij}(\kk_a+\qq)\PV^{jk}(\kk_b-\qq) \PV^{ki}(\qq) \nonumber \\
&&\hspace{10pt}
=2~\ccL^2~\ccV\left[\PP(k_a)\PP(k_b)\GG(\kk_a,\kk_b)+
\PP(k_b)\PP(k_c)\GG(\kk_b,\kk_c)+\PP(k_c)\PP(k_a)\GG(\kk_c,\kk_a)\right] 
 \\
&&\hspace{10pt}
+8~\ccV^3\int{d^3\qq\over(2\pi)^3}
\PP(|\kk_a+\qq|)\PP(|\kk_b-\qq|)\PP(q)
\GG(\kk_a+\qq,\kk_b-\qq)\GG(\kk_a+\qq,\qq)
\GG(\qq,\kk_b-\qq)
~,\nonumber
\enar
where the cross-power spectrum of the matter density and the relative
velocity is defined by
\beeq
\langle\delta_m(\kk_a)\UR^i(\kk_b)\rangle
=(2\pi)^3\delta^D(\kk_a+\kk_b)~\PX^i(\kk_a) 
=-\langle\UR^i(\kk_a)\delta_m(\kk_b)\rangle~
\eneq
as
\beeq
\PX^i(\kk)=-i~{k^i\over k^2}~{a\dot\TT_r(k)\over\SO}~\TT_m(k)\PPH(k) 
=i~{k^i\over k}~\TT_{ru}(k)\TT_m(k)\PPH(k)
=i~{k^i\over k}~{\TT_{ru}(k)\over\TT_m(k)}\PPM(k)~,
\label{eq:xxx}
\eneq
and the relative velocity kernel is
\beeq
\GG(\kk_a,\kk_b)=-{a^2\over\SO^2}~{\dot\TT_v(k_a)\over\TT_m(k_a)}
{\dot\TT_v(k_b)\over\TT_m(k_b)}~{\kk_a\cdot\kk_b\over k_a^2~k_b^2}
=-{\TT_{ru}(k_a)\over\TT_m(k_a)}~{\TT_{ru}(k_b)\over\TT_m(k_b)}~\mu_{ab}~,
\eneq
with $\mu_{ab}=\kk_a\cdot\kk_b/k_ak_b$. 

\begin{figure}
\centerline{\psfig{angle=-90, file=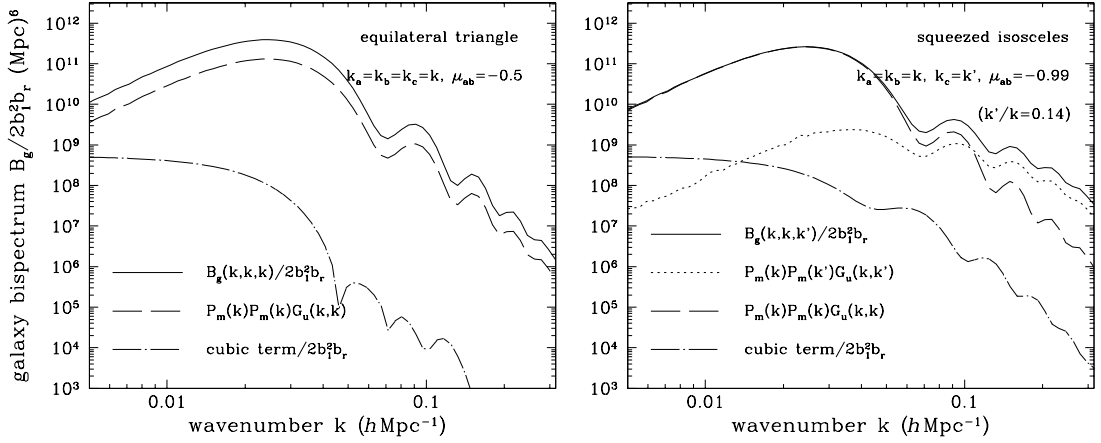, width=\textwidth}}
\caption{Relative velocity contributions to the galaxy bispectrum for an
equilateral (left) and a squeezed isosceles (right) triangular shapes.
Dotted, dashed, and dot-dashed lines show each component of the 
galaxy bispectrum
in eq.~(\ref{eq:bvl}), and their sum is shown as solid lines.
The bias parameter $\ccV/\ccL=0.01$ is assumed for computing the cubic term
in eq.~(\ref{eq:bvl}).}
\label{fig:f2}
\end{figure}

Figure~\ref{fig:f2} illustrates the galaxy bispectrum for two triangular
configurations. The left panel examines the scale-dependence of the galaxy
bispectrum for an equilateral shape ($k_a=k_b=k_c=k$, $\mu=-0.5$).
With the relative velocity kernel $\GG(k,k)\propto k^{-2}$ on scales
of interest, the galaxy bispectrum declines sharply on small scales
in proportion to $k^{-6.5}$, while it flattens on large scales.
However, because baryons and dark matter have similar velocities on
large scales $k\lesssim0.03\hmpci$ seen in Fig.~\ref{fig:f1}a, 
this term in the galaxy bispectrum
asymptotically vanishes. 
The acoustic oscillation structure of the galaxy bispectrum 
around $k\simeq0.1\hmpci$ reflects the oscillation due to
the relative velocity effect. Being a convolution, the cubic
contribution (dot-dashed) in eq.~(\ref{eq:bvl}) is constant and nonvanishing
on large scales, because the contribution to the cubic term arises
at $q\simeq0.01-0.1\hmpci$.
The right panel illustrates the scale-dependence
of the galaxy
 bispectrum for a squeezed isosceles triangle ($k_a=k_b=k\neq k_c$,
$\mu_{ab}=-0.99$). Since the third scale of the isosceles is larger
$k_c=k\sqrt{2+2\mu}=0.14~k$ and the scaling of the bispectrum is steep,
two contributions $\PPM(k)\PPM(k_c)\GG(k,k_c)$ in eq.~(\ref{eq:bvl})
that mix different scales $k$~and~$k_c$ are larger than the other term
$\PPM(k)\PPM(k)\GG(k,k)$ on small scales, but it falls over faster
on large scales.
The cubic term as in the equilateral configuration is subdominant
for $\ccV/\ccL\leq0.1$ at $k\geq0.01\hmpci$.

At low redshift, however, the matter fluctuation develops substantial
nonlinearity on small scales and is quasi-linear even at relatively large
scale $k\simeq0.1\hmpci$, demanding treatment beyond linear order
in eq.~(\ref{eq:basic}). The nonlinear evolution in the matter density
distribution results in a nonvanishing bispectrum of the galaxy fluctuation
even in the absence of the relative velocity effect. Here we adopt 
standard perturbation theory to compute the nonlinear terms in the matter 
density distribution to the third order
(e.g., \cite{GOGRET86,JABE94,MASASu92}):
$\delta_m(\kk)=\delta_m^\FI(\kk)+\delta_m^\SE(\kk)+\delta_m^\TI(\kk)$,
where the superscript indicates the perturbation order and
the second-order matter fluctuation is
\beeq
\delta_m^\SE(\kk)=\int{d^3\qq\over(2\pi)^3}~\delta_m^\FI(\qq)~
\delta_m^\FI(\kk-\qq)~\FF(\qq,\kk-\qq)~
\eneq
with the second-order kernel 
\beeq
\FF(\kk_a,\kk_b)={5\over7}+{2\over7}
\left({\kk_a\cdot\kk_b\over k_ak_b}\right)^2
+{\kk_a\cdot\kk_b\over2}\left({1\over k_a^2}+{1\over k_b^2}\right)~.
\eneq

\begin{figure}
\centerline{\psfig{angle=-90, file=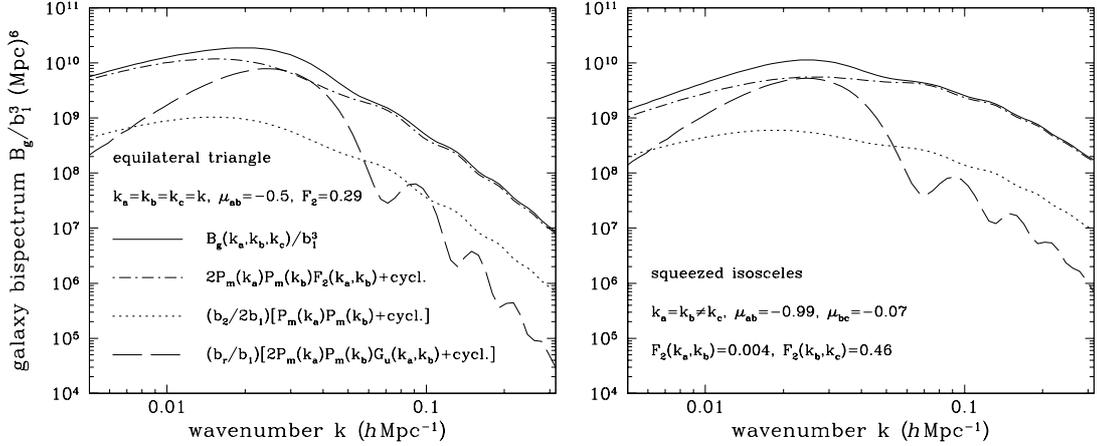, width=\textwidth}}
\caption{Scale-dependence of the full galaxy bispectrum for two triangular
shapes. Three different components contribute to the full galaxy bispectrum
in eq.~(\ref{eq:fbi}): The nonlinear evolution of the matter density
distribution (dot-dashed), the nonlinear bias (dotted), and the relative
velocity effect (dashed). The cubic term in eq.~(\ref{eq:fbi}) is omitted
to avoid clutter. The full galaxy bispectrum is shown as solid
lines. The bias parameters $\ccN/\ccL=0.1$ and $\ccV/\ccL=0.01$ are
assumed.}
\label{fig:f3}
\end{figure}

Along with the nonlinear gravitational evolution in the matter density
distribution, we also need to consider nonlinear galaxy bias.
Expanding to third order, we model the galaxy fluctuation as
\beeq
\label{eq:fullh}
\Dh(\xx)=\ccL~\delta_m(\xx)+{\ccN\over2}~\left[\delta_m^2(\xx)-
\sigma_m^2\right]
+{\ccH\over3!}~\delta_m^3(\xx)+\ccV\left[\UR^2(\xx)-\sur^2\right]~,
\eneq
where we keep only the linear order term in the relative velocity effect
as it is the relic effect of the early universe and it decays with the
expansion factor~$a$, rendering higher-order terms in $\UR^2(\xx)$
negligible at low redshift. Therefore,
the full bispectrum of the galaxy fluctuation in eq.~(\ref{eq:fullh}) is
\bear
\label{eq:fbi}
&&\BB(\kk_a,\kk_b,\kk_c)=\ccL^3~\left[2\PP(k_a)\PP(k_b)\FF(\kk_a,\kk_b)
+\up{cycl.}\right]
+{1\over2}~\ccL^2~\ccN\left[\PP(k_a)\PP(k_b)
+\up{cycl.}\right]  \\
&&+\ccL^2\ccV\left[2\PP(k_a)\PP(k_b)\GG(\kk_a,\kk_b)
+\up{cycl.}\right] \nonumber \\
&&+8~\ccV^3\int{d^3\qq\over(2\pi)^3}
\PP(|\kk_a+\qq|)\PP(|\kk_b-\qq|)\PP(q)
\GG(\kk_a+\qq,\kk_b-\qq)\GG(\kk_a+\qq,\qq)\GG(\qq,\kk_b-\qq)
~, \nonumber 
\enar
where $\PP(k)$ is the linear matter power spectrum.
In addition to the
relative velocity contributions in eq.~(\ref{eq:bvl}),
the full galaxy bispectrum receives two more contributions from the nonlinear
evolution: The first square-bracket
represents the contributions of the nonlinear evolution in the matter 
density distribution, while the second square-bracket represents the nonlinear
bias contributions described in eq.~(\ref{eq:fullh}).

Figure~\ref{fig:f3} plots the scale-dependence of the full galaxy bispectrum
in eq.~(\ref{eq:fbi}). The left panel dissects each component of the galaxy 
bispectrum given an equilateral triangular shape ($k_a=k_b=k_c=k$, $\mu=-0.5$,
$\FF=0.29$).  In this and subsequent figures,
we assume the nonlinear bias parameter $\ccN/\ccL=0.1$ and
the relative velocity bias parameter $\ccV/\ccL=0.01$ as our fiducial
bias parameters for illustration.
Two contributions from the nonlinear bias (dotted)
and matter density (dot-dashed) are comparable in amplitude 
if $\ccN\simeq\ccL$, and they are dominant over 
the relative velocity effect (dashed) on small scales.
However, the relative velocity contribution is comparable to the matter
density contribution at large scale $k\simeq0.03\hmpci$, corresponding to the
peak seen in Fig.~\ref{fig:f1}a.
The right panel shows the galaxy bispectrum for a squeezed
isosceles shape ($k_a=k_b=k\neq k_c$, $\mu_{ab}=-0.99$). For the squeezed
configuration, the large scale power $\PPM(k_c)$ with $k_c=0.14k$
enhances all three components on small scales and reduces the components
on large scales. The relative velocity peak leaves a bump at 
$k\simeq0.02\hmpci$ in the squeezed configuration, 
similar to the scale seen in
the equilateral case. Therefore, the unique signature of the
relative velocity effect in the galaxy bispectrum on large scales
can be used to robustly measure the relative velocity effect 
for $\ccV\geq0.01$ in low-redshift massive galaxies.

\section{Effect on BAO measurements: Power spectrum}
\label{sec:power}

In addition to generating a large-scale bispectrum, the relative
velocity effect can also modify the galaxy power spectrum on scales of order
the sound horizon.  Numerous ongoing and future galaxy surveys seek to
measure the galaxy power spectrum on these scales with extraordinary
precision to localize the BAO feature and thereby reconstruct the
expansion history of the universe, constraining the kinematic
properties of dark energy.  From the viewpoint of BAO probes of dark
energy, the relative velocity effect could be a significant
contaminant, as we illustrate in this section.  

\subsection{Relative velocity contribution and galaxy power
spectrum}
\label{ssec:pow}

\begin{figure}
\centerline{\psfig{file=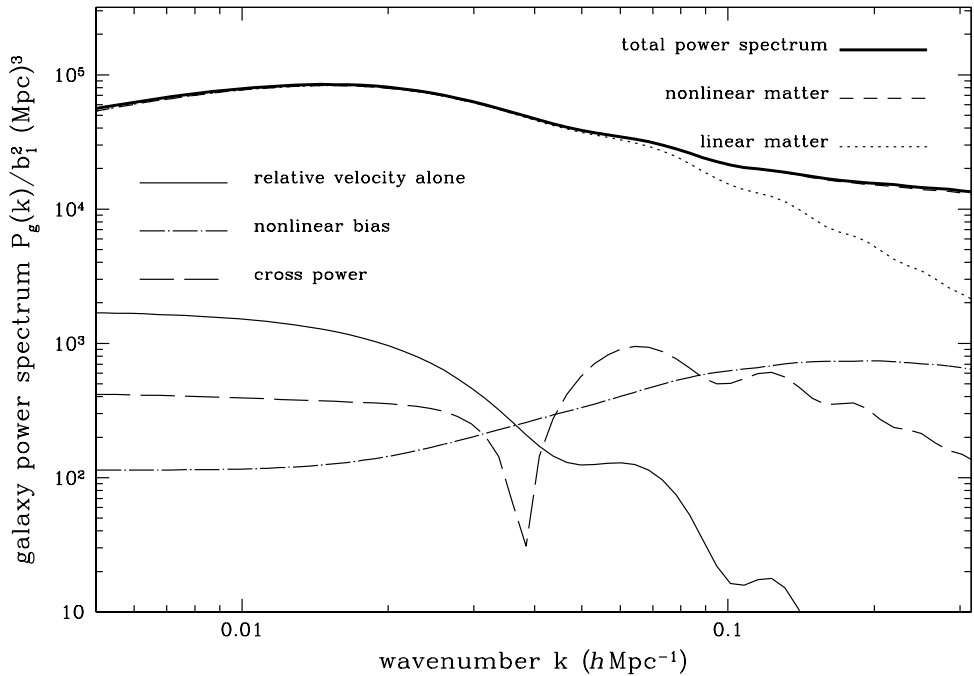, width=5in}}
\caption{Full galaxy power spectrum with the bias parameters
$\ccN/\ccL=0.1$ and $\ccV/\ccL=0.01$.
The galaxy power spectrum (thick solid) is computed by using 
eq.~(\ref{eq:fullp}).
Thin solid, long dashed, and dot-dashed lines represent the auto-power
spectrum of the relative velocity effect ($\sim\ccV^2$), the cross-power
spectrum of the relative velocity and the nonlinear galaxy bias 
($\sim\ccV$), and
the auto-power spectrum of the nonlinear galaxy bias, respectively.
The latter (long dashed) becomes negative at $k=0.04\hmpci$, beyond which 
its absolute value is plotted.
The linear and the nonlinear matter power spectra are shown as dotted and
short dashed lines (the short dashed line is largely obscured by the thick
solid line).}
\label{fig:f4}
\end{figure}

We compute the power spectrum of the galaxy fluctuation using 
eq.~(\ref{eq:fullh}), but for simplicity we first compute the relative velocity
contributions only. With caution that the cross-power spectrum $\PX^i(k)$ in 
eq.~(\ref{eq:xxx}) depends on the direction of the wavevector,
the ensemble average of the galaxy fluctuation is
\bear
\langle\Dh(\kk_a)\Dh(\kk_b))\rangle&=&\ccV
\int{d^3\qq_1\over(2\pi)^3}\int{d^3\qq_2\over(2\pi)^3}\sum_{i=x,y,z}
\bigg[2~\ccL\FF(\qq_1,\kk_a-\qq_1)
\left\langle\delta_m(\qq_1)\delta_m(\kk_a-\qq_1)\UR^i(\qq_2)\UR^i(\kk_b-\qq)
\right\rangle  \nonumber\\
&&\hspace{20pt}+\ccN\left\langle\delta_m(\qq_1)\delta_m(\kk_a-\qq_1)
\UR^i(\qq_2)\UR^i(\kk_b-\qq)\right\rangle\bigg] \nonumber \\
&&+\ccV^2\int{d^3\qq_1\over(2\pi)^3}\int{d^3\qq_2\over(2\pi)^3} 
\sum_{i,j=x,y,z}\langle\UR^i(\qq_1)\UR^i(\kk_a-\qq_1)
\UR^j(\qq_2)\UR^j(\kk_b-\qq)\rangle~, \nonumber
\enar
and by isolating the connected part only the galaxy power spectrum can
be computed as
\bear
\label{eq:relp}
\Ph(\kk)&=&\ccV\int{d^3\qq\over(2\pi)^3}\!\!\sum_{i=x,y,z}\!\!\!\!
\PX^i(\qq)\PX^i(\kk-\qq)\bigg[4~\ccL\FF(\qq,\kk-\qq)+2~\ccN\bigg]
+2~\ccV^2\int{d^3\qq\over(2\pi)^3}\!\!\sum_{i,j=x,y,z}\!\!\!\!
\PV^{ij}(\qq)~\PV^{ij}(\kk-\qq) \nonumber \\
&=&\int{d^3\qq\over(2\pi)^3}\PP(q)\PP(|\kk-\qq|)\GG(\qq,\kk-\qq)
\bigg[4~\ccL\ccV\FF(\qq,\kk-\qq)
+2~\ccN\ccV+2~\ccV^2\GG(\qq,\kk-\qq)\bigg]~
\enar
(see Appendix~\ref{ap:com} for derivation).
The third term in the square-bracket is the pure relative velocity contribution
to the galaxy power spectrum, while the other terms represent
the contributions of the relative velocity effect in conjunction with 
the nonlinear galaxy bias.

Finally, accounting for all the remaining contributions from the nonlinear
evolution in the matter density distribution, 
the full power spectrum of the galaxy fluctuation can be written as
\bear
\label{eq:fullp}
\Ph(\kk)&=&\ccL^2~\PPN(\kk)+\int{d^3\qq\over(2\pi)^3}\PP(q)\PP(|\kk-\qq|)
\bigg[{1\over2}~\ccN^2+2~\ccL\ccN\FF(\qq,\kk-\qq) \nonumber \\
&&+4~\ccL\ccV\FF(\qq,\kk-\qq)\GG(\qq,\kk-\qq)
+2~\ccN\ccV\GG(\qq,\kk-\qq)
+2~\ccV^2\GG(\qq,\kk-\qq)^2\bigg] ~.
\enar
To second order in the power spectrum, the linear bias parameter~$\ccL$ is 
renormalized.  $\PPN(k)$ in the first term is the nonlinear matter power
spectrum, while $\PP(k)$ in the integral is computed by using the linear
matter power spectrum \cite{HEMAVE98,MCDON06}.
Compared to eq.~(\ref{eq:relp}), two 
additional contributions in the square-bracket arise from the 
nonlinear galaxy bias.

Figure~\ref{fig:f4} shows the full galaxy power spectrum computed by using
eq.~(\ref{eq:fullp}). Assuming our fiducial bias parameters
$\ccN/\ccL=0.1$ and $\ccV/\ccL=0.01$, the galaxy power spectrum (thick solid)
is largely determined by the matter power spectrum (linear: dotted,
nonlinear: short dashed) on all scales, yet the relative velocity effect
and the nonlinear galaxy
bias affect the galaxy power spectrum at the percent level
on various scales. These contributions expressed in the second
square-bracket of eq.~(\ref{eq:fullp}) are split into three components 
with different dependences on the relative velocity bias~$\ccV$: The
auto-power spectrum of the relative velocity effect ($\ccV^2$: thin
solid), the cross-power spectrum of the relative velocity and the
nonlinear galaxy bias ($\ccV$: long dashed), and the
auto-power spectrum of the nonlinear galaxy bias (dot-dashed).
The auto-power spectrum (thin solid) of the relative velocity effect
closely resembles the velocity power spectrum with the prominent
oscillation structure seen in Fig.~\ref{fig:f1}a.
The auto-power spectrum (dot-dashed) of the 
nonlinear galaxy bias is constant on large scales and approaches the
shape of the matter power spectrum on small scales, while its power is enhanced
at $k\simeq0.2\hmpci$ by the nonlinear kernel~$\FF(\qq,\kk-\qq)$ that
puts more weight on the large-scale power in the convolution. The cross-power
spectrum (long dashed) of the relative velocity and the nonlinear galaxy
bias changes
sign at $k\simeq0.04\hmpci$, reflecting the out-of-phase nature of the
velocity and the matter distributions seen in Fig.~\ref{fig:f1}b.

Future galaxy surveys that use BAO probes of dark energy will measure the
large-scale galaxy power spectrum and attempt to determine the BAO
peak position at sub-percent level precision.  At $k\simeq0.1\hmpci$,
the auto-power spectrum of the relative velocity effect (thin solid)
is rather sub-dominant compared to the contribution of the cross term
and the nonlinear galaxy bias term (long dashed and dot-dashed) for our
fiducial parameters.  Therefore, both the nonlinear bias~$\ccN$ 
and the relative velocity bias~$\ccV$ parameters must be
determined well in order to completely clean out the relative velocity
effect in the galaxy power spectrum.  This trend remains unchanged for
$\ccV/\ccL<0.1$ given $\ccN/\ccL=0.1$.

\begin{figure}
\centerline{\psfig{file=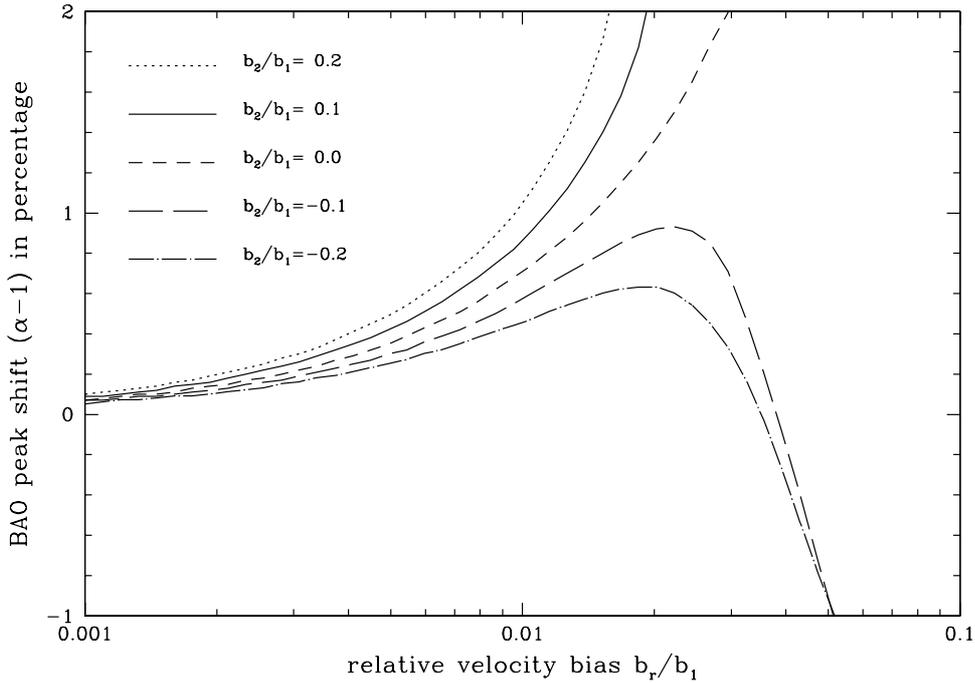, width=5in}}
\caption{Impact of the relative velocity effect on the BAO peak shift.
Various lines represent the peak shift $(\alpha-1)$ in percentage 
as a function of the relative velocity
bias parameter $\ccV/\ccL$, given the value of the
nonlinear galaxy bias parameter $\ccN/\ccL$.
Following the \citet{SESIET08} approach,
the peak shift is obtained by fitting the template power spectrum to
the full galaxy power spectrum in eq.~(\ref{eq:fullp}) accounting for the
nonlinear growth and the anomalous power (see text for detail).}
\label{fig:f5}
\end{figure}

\subsection{Shift in the BAO peak position}
\label{ssec:bao}
We quantify the impact of the relative velocity effect on the precision 
measurement of the baryonic acoustic oscillation feature in future
dark energy surveys, assuming that {\it no}
 information is utilized for cleaning
out the relative velocity effect in the power spectrum measurement.
Following the \citet{SESIET08} approach to modeling the scale-dependent
nonlinear growth and anomalous power in the matter power spectrum, we fit
the template power spectrum to the full galaxy power spectrum computed in
eq.~(\ref{eq:fullp}) to characterize the BAO peak shift.
The template power spectrum is parametrized as
\beeq
\PPT(k)=(c_0+c_1k+c_2k^2)P_\up{evo}(k/\alpha)+(a_0+a_1k+a_2k^2+\cdots+a_7k^7)
\label{eq:temp}
\eneq
with polynomial coefficients $c_i$ and $a_i$ taken as free parameters.
Any deviation of the parameter~$\alpha$ from unity represents the BAO peak
shift in measurements.
The two groups of polynomials in $k$ (multiplicative and additive)
account for the scale-dependent
nonlinear growth and the additive broad-band power; This parametrization
corresponds to the ``Poly7'' fit in \citet{SESIET08}.
The evolved linear matter power spectrum is then given in the form
\beeq
P_\up{evo}(k)=\left[P_\up{lin}(k)-P_\up{no-wiggle}(k)\right]
\exp\left(-k^2\Sigma_m^2/2\right)+P_\up{no-wiggle}(k)~,
\eneq
where $P_\up{no-wiggle}(k)$ is the fit to the smooth 
power spectrum shape without
the BAO wiggle from \citet{EIHU98} and $\Sigma_m=8.8~\hmpc$ accounts for the
degradation of the BAO wiggle in time ($z=0$) \cite{SESIET08}.

In Fig.~\ref{fig:f5} we perform a $\chi^2$ analysis to compute the BAO peak
shift as a function of the bias parameters $\ccV/\ccL$ and $\ccN/\ccL$
by fitting the template power spectrum in eq.~(\ref{eq:temp}) to the
full galaxy power spectrum in eq.~(\ref{eq:fullp}) over a range of
$0.02\hmpci<k<0.35\hmpci$. Since nonlinear effects do shift the BAO
peak position ($\sim0.5\%$ at $z=0.3$ \cite{SESIET08})
at low redshift,
we isolate the effect of the relative velocity
on the BAO peak shift by setting $\Delta\alpha=0$
when $\ccV/\ccL=0$. 
The relative velocity effect shifts the BAO peak position no more than
1\% at $\ccV/\ccL\leq0.01$ with a reasonable range of the nonlinear galaxy bias
parameter, but its impact increases dramatically at $\ccV/\ccL\gg0.01$,
because the acoustic structure in the relative velocity effect is 
anti-correlated with the structure in the matter distribution. While a
large nonlinear galaxy 
bias can affect the BAO peak position, its parameters are, by contrast, 
relatively well constrained, and its impact is at the sub-percent level
for our fiducial value $\ccV/\ccL$ 
over a range of $\ccN/\ccL$ values considered
here. For negative values of $\ccN/\ccL$, the cross-power spectrum 
(long dashed in Fig.~\ref{fig:f4}) of the nonlinear galaxy
bias becomes positive, but the auto-power spectrum (dot-dashed in
Fig.~\ref{fig:f4}) of the nonlinear galaxy bias is reduced due to the sign
change in $\ccN$. At $k\geq0.1\hmpci$, the latter is dominant over the
cross-power spectrum, and hence the peak shift is reduced at $\ccV/\ccL=0.01$,
compared to the case with $\ccN/\ccL>0$. However, as the relative velocity
bias increases, the cross-power spectrum contributes more to the peak shift,
changing its direction. The auto-power spectrum 
(thin solid in Fig.~\ref{fig:f4}) of the pure relative velocity contribution
around the BAO scale is rather sub-dominant over the scales for the range
of $\ccV/\ccL$ considered here.

\section{Cosmological constraining power and bias in parameter estimation}
\label{sec:for}
In section~\ref{sec:power} we showed that the presence of the relative 
velocity effect can adversely affect our ability to measure the BAO peak
position and constrain the cosmological parameters when its effect is ignored
in the power spectrum analysis. However, it was also demonstrated in 
section~\ref{sec:bispectrum} that measurement of the bispectrum provides
a promising way to measure the relative velocity effect and account for it
when we estimate the cosmological parameters. In this section,
we perform a Fisher matrix analysis to answer three questions: What is the
bias in our parameter estimation if the relative velocity effect is unaccounted
for? How well can we measure or constrain the relative velocity effect
using the galaxy power spectrum and bispectrum measurements? And how much would
the cosmological parameter constraints be inflated if we included the relative
velocity bias as additional parameter? To answer these questions and
for definiteness,
we consider a galaxy survey of volume $V=10~(\hgpc)^3$ with number density
$n_g=10^{-4}(\hmpc)^3$. As our cosmological model, we assume a flat
universe with a constant dark energy equation-of-state and add
two bias parameters $\ccN/\ccL$ and $\ccV/\ccL$ to describe the
nonlinear galaxy bias and the relative velocity distribution:
$\pp=(n_s,\alpha_s,\omega_m,\omega_b,\omega_\up{de},w_0,\ccN/\ccL,\ccV/\ccL)$ 
with $h^2=\omega_m+\omega_\up{de}$ (see section~\ref{sec:intro} for
our choice of the fiducial model parameters).\footnote{We assume that the
linear bias~$\ccL$ and the matter fluctuation normalization $\Delta^2_\varphi$
is degenerate in the power spectrum analysis, and we can only constrain its
combination. The degeneracy is partially broken by the nonlinear effect
and also by the bispectrum, but since the normalization is nuisance in the
present analysis we combine it with the linear bias term and remove it from
our parameter set by assuming that other parameters are not affected by
the change in the overall normalization of the measurements.}

First, we compute the impact of the relative velocity effect on cosmological
parameter estimation from galaxy power spectrum measurements when the relative
velocity effect is unaccounted. This case will represent the currently
planned galaxy surveys, as the null hypothesis is 
devoid of the relative velocity
effect. Here we quantify its impact in terms of the parameter
bias $\Delta w_0$ in the dark energy equation-of-state.
Our model parameters are described by 
$\bar\pp\equiv\pp\left[\ccV/\ccL=0\right]$, 
in which the relative velocity bias parameter is {\it incorrectly} set zero
($\ccV=0$), while in reality it is nonzero ($\ccV\neq0$).
The parameter bias in our estimation of the dark energy equation-of-state is 
then (see Appendix~\ref{ap:bias})
\beeq
\label{eq:wbias}
\Delta w_0=\sum_i^{\bar\pp}\bigg[F^{-1}(\bar\pp)\bigg]_{w_0i}\times
\sum_{k=k_\up{min}}^{k_\up{max}}
{1\over\sigma^2_{\Ph}(k)}{\partial\Ph(k)\over\partial p_i}
\bigg[\Ph(k|\pp)-\Ph(k|\bar\pp)\bigg]~,
\eneq
where the Fisher information matrix is
\beeq
\label{eq:fisher}
F_{ij}(\bar\pp)=\sum_{k=k_\up{min}}^{k_\up{max}}
{1\over\sigma^2_{\Ph}(k)}
{\partial\Ph(k|\bar\pp)\over\partial p_i}
{\partial\Ph(k|\bar\pp)\over\partial p_j}~,
\eneq
the uncertainty in the power spectrum measurements is \cite{FEKAPE94}
\beeq
\sigma_{\Ph}^2(k)={2\over4\pi^2k^2\Delta k ~V/(2\pi)^3}
\left[\Ph(k)+{1\over n_g}\right]^2~,
\eneq
and $k_\up{min}=\Delta k=2\pi/V^{1/3}=0.0029~\hmpci$. In computing the
Fisher information matrix in eq.~(\ref{eq:fisher}) we also add the Planck
prior on the cosmological parameters, following the procedure described
in the Appendix of the Dark Energy Task Force final report
\cite{DETF06} (see also \cite{ZASPSE97,BOEFTE97}).

\begin{figure}
\centerline{\psfig{angle=-90, file=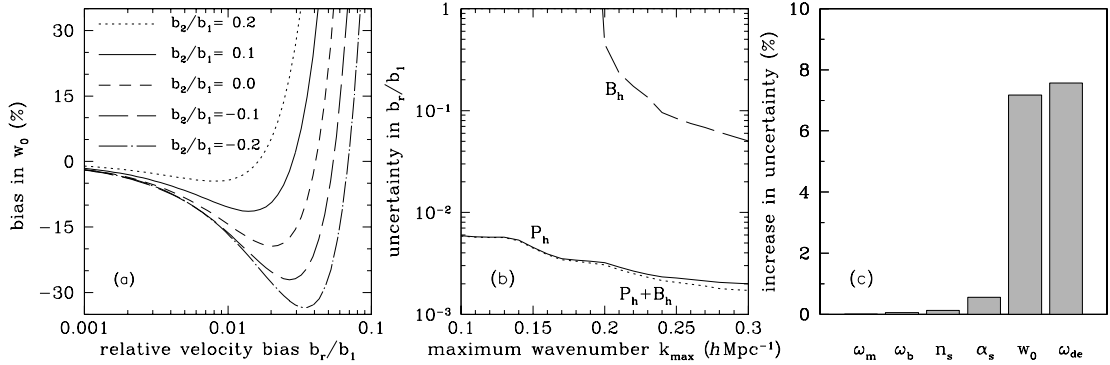, width=\textwidth}}
\caption{Impact of the relative velocity effect on the cosmological parameter
estimation. For computing the constraint on the dark energy 
equation-of-state $w_0$, we use the galaxy power spectrum measurements in 
a  survey of volume $V=10~(\hgpc)^3$ with 
number density $n_g=10^{-4}(\hmpc)^{-3}$.
Parameter constraints are obtained by adding CMB priors from Planck
and marginalizing over the remaining
parameters (see text for details). ($a$)~Parameter bias 
in measuring $w_0$ 
when the relative velocity effect is unaccounted. 
($b$)~Measurement uncertainty in the relative velocity bias parameter 
(the nonlinear galaxy bias parameter is fixed $\ccN/\ccL=0.1$).
($c$)~Fractional increase in the cosmological parameter constraints when the 
relative velocity bias is modeled and marginalized over 
($k_\up{max}=0.2\hmpci$).}
\label{fig:f6}
\end{figure}

Figure~\ref{fig:f6}a shows the bias in the dark energy
equation-of-state $w_0$ as a function of the relative velocity bias
given various values of the nonlinear galaxy bias $\ccN/\ccL$. We have fixed
$k_\up{max}=0.2\hmpci$.
Naturally, the bias $\Delta w_0$ increases with larger value of $\ccV/\ccL$,
as the relative velocity effect is unaccounted for.
Negative values of $\ccN/\ccL$ reduce the auto-power spectrum (dot-dashed
in Fig.~\ref{fig:f4}) of the nonlinear galaxy bias and change the sign
of its cross-power spectrum (long dashed in Fig.~\ref{fig:f4}) with the
relative velocity on small scales, enhancing the difference in the power
spectrum around the BAO scale
and thereby increasing the bias $\Delta w_0$, compared to reducing
the relative velocity contribution in the fiducial model 
with positive $\ccN/\ccL$. Meanwhile, since
our parameter estimation is affected by
the power spectrum measurements on all scales rather than just on BAO
scales,  there is a cross-over point
in $\ccV/\ccL$, at which the bias from the cross-power spectrum is
cancelled by the bias from the auto-power spectrum of the relative velocity.
Moreover, the parameter bias obtained here is rather
insensitive to the precision of the power spectrum measurements, since it
arises from our {\it incorrect}
 modeling of the mean value of the power spectrum.
For the fiducial value $\ccN/\ccL=0.1$ (solid), the bias $\Delta w_0$ is
at the 10\% level for $\ccV/\ccL=0.01$ and 2\% for $\ccV/\ccL=0.001$.
While a more thorough analysis of the BAO peak position may further reduce
the parameter bias than our simple estimates using the Fisher matrix
(like in section~\ref{ssec:bao}, e.g., \cite{SESIET08,PAWH08}),
Fig.~\ref{fig:f6}a demonstrates the significance of the relative velocity
effect on the future dark energy surveys, and the relative velocity bias
parameter
should be included and marginalized over in the power spectrum analysis.

Next, we estimate our ability to measure the relative velocity effect
from the galaxy
power spectrum and bispectrum measurements. Our fiducial model
in this case is then described by $\pp[\ccV/\ccL=0]$, correctly
representing the reality. The measurement uncertainty in the
 relative velocity bias parameter can be expressed as
\beeq
\sigma^2_{\ccV/\ccL}=\left[F^{-1}(\pp)\right]_{\ccV/\ccL~\ccV/\ccL}~,
\eneq
and the Fisher matrix~$F_{ij}$ now includes that of the bispectrum
when the bispectrum measurements are used \cite{SCSEZA04,SECRET06,BASESE10},
\beeq
F_{ij}=
\sum_{k_a=k_\up{min}}^{k_\up{max}}
\sum_{k_b=k_\up{min}}^{k_a}
\sum_{k_c=k^\star_\up{min}}^{k_b}
{1\over\sigma^2_{\BB}}
{\partial\BB(k_a,k_b,k_c)\over\partial p_i}
{\partial\BB(k_a,k_b,k_c)\over\partial p_j}~
\eneq
with the bispectrum variance
\beeq
\sigma_{\BB}^2(k_a,k_b,k_c)={s_{abc}\over8\pi^2k_ak_bk_c\Delta k^3
 ~V/(2\pi)^3}
\left[\Ph(k_a)+{1\over n_g}\right]
\left[\Ph(k_b)+{1\over n_g}\right]
\left[\Ph(k_c)+{1\over n_g}\right]~,
\eneq
where $k^\star_\up{min}=\up{max}(k_\up{min},|k_a-k_b|)$ and the symmetry
factor is $s_{abc}=6,2,1$ for equilateral, isosceles, and general triangular
configurations, respectively. We have assumed that the bispectrum estimates
are Gaussian distributed.

Figure~\ref{fig:f6}b shows the measurement uncertainty
$\sigma_{\ccV/\ccL}$ in the relative velocity bias parameter,
after all the remaining parameters~$\pp$ are marginalized
over, which can be translated into the detection significance
$S/N\simeq(\ccV/\ccL)/\sigma_{(\ccV/\ccL)}$~. With the few percent level
contribution of the relative velocity effect at $\ccV/\ccL=0.01$ around the
BAO scale seen in Fig.~\ref{fig:f4}, power spectrum measurements (solid)
alone can constrain the relative velocity bias as small as
$\ccV/\ccL=0.01$ at the $2-3\sigma$ confidence level.
 Bispectrum measurements (dashed) provide additional
but less stringent constraint on the relative velocity bias parameter
if $\ccV=0$, than power spectrum measurements,
especially when the basic cosmological parameters are already well constrained
by CMB measurements. Hence, the combination (dotted) of the power
spectrum and bispectrum measurements yields constraints mainly derived
from the power spectrum measurements. However, as illustrated in 
Fig.~\ref{fig:f3}, the bispectrum exhibits a unique signature of the relative
velocity effect on large scales, if the relative velocity effect is
{\it present}. Therefore,
if any hint of the relative velocity effect is found in the
power spectrum analysis, measurements of the bispectrum can be used to confirm
and constrain the relative velocity effect in an indisputable way.

Finally, we investigate the {\it cost} in the parameter constraints by adding 
the additional parameter $\ccV/\ccL$ and marginalizing over it. 
Figure~\ref{fig:f6}c compares the parameter constraints in two cases,
when the relative
velocity bias parameter is modeled, and when it is simply assumed zero,
i.e., $\sigma_{p_\alpha}(\pp)/\sigma_{p_\alpha}(\bar\pp)$~. 
The matter and the baryon density
constraints $\sigma_{\omega_m}$ and $\sigma_{\omega_b}$ are little affected by the
addition of the relative velocity bias parameter, as they are largely
independent. The constraints on the spectral
index $n_s$ and its running $\alpha_s$ become weaker by
 a few percent level, compared to the constraints in the model
without the relative velocity bias parameter, indicating that the relative
velocity effect is somewhat degenerate with~$n_s$ and~$\alpha_s$.
Last, as we showed in previous sections the relative velocity effect can
adversely affect the dark energy parameter estimation, and Fig.~\ref{fig:f6}c
shows that the constraints on dark energy
parameters $w_0$ and $\omega_\up{de}$ would degrade by $\sim8\%$,
if we included the relative velocity bias parameter.
While the spectral index and its running are also affected by the relative
velocity parameter, they are highly constrained by CMB measurements,
 in contrast
to the dark energy parameters. Meanwhile, the inflated constraints on the
dark energy parameters can be reduced by using the bispectrum 
measurement to improve the constraint on the relative velocity bias parameter
and thereby recovering the dark energy constraints.
Overall, the cost of modeling the
relative velocity bias is rather low, especially considering the magnitude of
the parameter bias when the relative velocity effect is unaccounted for.

\section{Discussion}
\label{sec:discussion}

We have investigated the relative velocity effect of the baryon and the
dark matter distributions on the baryonic acoustic oscillation (BAO) 
measurements at low redshift. The relative velocity effect imprinted in
the earliest baryonic structure at high redshift leaves its signature
in the power spectrum of these objects \cite{DAPESE10}, which in turn
alters the subsequent star formation and reionization history, affecting
the large-scale clustering of massive galaxies at low redshift. The relative
velocity effect that persists until the late time results in a nonvanishing
galaxy bispectrum, and its contribution to the bispectrum can be easily
isolated
on large scales $k\simeq0.03\hmpci$, where the relative velocity contribution
peaks.
Since the acoustic structure of the relative velocity effect
is non-smooth and out-of-phase compared to the structure in the matter 
distribution, its effect cannot be removed by using a broad-band filter that 
is designed to deal with the nonlinear matter evolution, and the relative
velocity effect can potentially bias the BAO peak position determination.

We find that the relative velocity effect can, if unaccounted, 
shift the BAO peak position measurements and bias the estimates
of the dark energy equation-of-state by $\sim$10\%, depending on
the amplitude of the relative velocity effect. However, the power spectrum
measurement itself is sensitive enough to the relative velocity effect that
we can measure its amplitudes as small as $\ccV/\ccL\simeq0.01$
along with other cosmological parameters. Including the relative velocity
bias as additional parameter may inflate the constraint $\sigma_{w_0}$
on the dark energy equation-of-state by 8\%, a marginal cost
to pay, compared to the potentially pernicious risk arising from the 
unaccounted relative
velocity effect. Bispectrum measurements also provide
tight constraints on the relative velocity effect, though its constraint 
in the fiducial model without the relative velocity effect is less stringent
than from power spectrum measurements due to larger measurement uncertainties.
However, the bispectrum signal is substantially enhanced, especially on large
scales, if the relative velocity effect is present, providing alternative way
 to confirm and measure the relative velocity effect in a more robust
and model-independent way.

Many future dark energy surveys such as 
BigBoss,\footnote{http://bigboss.lbl.gov}
Euclid,\footnote{http://sci.esa.int/euclid}
and the Wide-Field Infrared Survey
Telescope\footnote{http://wfirst.gsfc.nasa.gov}
 plan on measuring galaxies at moderately high redshift, $z\gtrsim 1$,
where nonlinear growth is less severe and even smaller scales
may safely be included in the analysis. 
However, these high redshift surveys are
not immune to the relative velocity effect, which grows in amplitude
like $(1+z)$, eventually dominating over the matter 
fluctuations at $z\gg10$. At the redshifts probed by these BAO surveys,
the largest uncertainty in our calculation is the amplitude of the relative
velocity bias parameter $\ccV$. Any effect of relative velocities
on massive galaxies at such redshifts must be indirect, making
difficult any theoretical estimation of the relative velocity bias parameter.
It is quite plausible that the complicated process of massive
galaxy formation obliterates the relative velocity effect
at low redshift, meaning that this effect may not be apparent in the
distribution of low-redshift massive galaxies.
Nevertheless, we note that the relative velocity effect can be
easily modeled without significantly
inflating the cosmological parameter constraints,
and if present but unaccounted, it can catastrophically affect the
ability of BAO survey to perform high
precision cosmological parameter estimation.

\acknowledgments
We thank Tobias Baldauf, Chris Hirata, Hee-Jong Seo,
and Dmitriy Tseliakhovich for useful discussions.
J.Y. is supported by the SNF Ambizione Grant.
This work is supported by the Swiss National Foundation under contract 
200021-116696/1 and WCU grant R32-10130.

\bibliography{sonic.bbl}

\appendix
\section{Power spectrum and bispectrum computation}
\label{ap:com}
Here we derive key equations for computing the full galaxy power
spectrum in eq.~(\ref{eq:fullp}) and bispectrum in eq.~(\ref{eq:fbi}).
Full computation of the galaxy power spectrum requires a volume integration
over the wavevector~$\qq$, and it can be obtained by a combination of the
two integrations:
\bear
&&\int{d^3\qq\over(2\pi)^3}\PP(q)\PP(|\kk-\qq|)\FF(\qq,\kk-\qq)\\
&&\hspace{30pt}
={k^3\over2\pi^2}\int_0^\infty dr~r^2\PP(kr)\int_{-1}^1d\mu~
\PP\left(k\sqrt{1+r^2-2r\mu}\right)
\left[{3r+7\mu-10r\mu^2\over14~r(1+r^2-2r\mu)}\right]~,
\nonumber    \\
&&\int{d^3\qq\over(2\pi)^3}\PP(q)\PP(|\kk-\qq|)\GG(\qq,\kk-\qq)\\
&&\hspace{10pt}
={k^3\over2\pi^2}\int_0^\infty dr~r^2\PP(kr)\int_{-1}^1d\mu~
\PP\left(k\sqrt{1+r^2-2r\mu}\right)
\left[{\TT_{ru}(q)\over\TT_m(q)}{\TT_{ru}(k\sqrt{1+r^2-2r\mu})
\over\TT_m(k\sqrt{1+r^2-2r\mu})}
{r-\mu\over\sqrt{1+r^2-2r\mu}}\right]~,
\nonumber
\enar
where $r=q/k$ is the ratio of the two wavevectors $\kk$ and $\qq$ and
$\mu$ is their cosine angle.

For the bispectrum computation, since the connected wavevectors
($\kk_a+\kk_b+\kk_c=0$) defines a 2D surface rather than a line
($\kk_a+\kk_b=0$) in the power spectrum case,
the azimuthal symmetry in the volume integration is broken, and a full 3D
integration needs to be performed. Given a triangular configuration
($k_a$, $k_b$, and $\mu_{ab}=\kk_a\cdot\kk_b/k_ak_b$), the last term of
the full bispectrum in eq.~(\ref{eq:fbi}) is
\bear
&&\int{d^3\qq\over(2\pi)^3}
\PP(|\kk_a+\qq|)\PP(|\kk_b-\qq|)\PP(q)
\GG(\kk_a+\qq,\kk_b-\qq)\GG(\kk_a+\qq,\qq)\GG(\qq,\kk_b-\qq) \\
&&\hspace{30pt}
=-{k^3\over2\pi^2}\int_0^\infty dr~r^2\PP(kr)
{\TT^2_{ru}(kr)\over\TT^2_m(kr)}
\int_{-1}^1{d\mu\over2}~
\PP\left(k_a\sqrt{1+r^2+2r\mu}\right)
{\TT^2_{ru}(k_a\sqrt{1+r^2+2r\mu})\over\TT^2_m(k_a\sqrt{1+r^2+2r\mu})}
\left[{\mu+r\over \sqrt{1+r^2+2r\mu}}\right]
\nonumber\\
&&\hspace{30pt}\times
\int_0^{2\pi}{d\phi\over2\pi}
\PP\left(k_a\sqrt{r^2+r'^2-2rr'\mu_{bq}}\right)
{\TT^2_{ru}\left(k_a\sqrt{r^2+r'^2-2rr'\mu_{bq}}\right)
\over\TT^2_m\left(k_a\sqrt{r^2+r'^2-2rr'\mu_{bq}}\right)} 
\nonumber \\
&&\hspace{70pt}\times
{r'\mu_{ab}-r\mu+rr'\mu_{bq}-r^2\over \sqrt{1+r^2+2r\mu}
\sqrt{r^2+r'^2-2rr'\mu_{bq}}}
{r'\mu_{bq}-r\over\sqrt{r^2+r'^2-2rr'\mu_{bq}}}~,
\nonumber
\enar
with $r'=k_b/k_a$ and
$\mu_{bq}=\mu_{ab}\mu+\sqrt{1-\mu_{ab}^2}\sqrt{1-\mu^2}\cos\phi$~,
where $\phi$ is the azimuthal angle of $\qq$.

\section{Parameter forecast and bias in the parameter estimation}
\label{ap:bias}
Assuming the measurements of the galaxy power spectrum and bispectrum are
Gaussian distributed, we can compute the likelihood of the measurements 
given a set of model parameters~$\pp$,
\beeq
2\mathcal{L}(\pp)=\ln\det\CC+(\xx-\mm)^T\CC^{-1}(\xx-\mm)+\up{constant} ~,
\label{eq:like}
\eneq
where~$\xx$ and~$\mm(\pp)$ respectively represent the measurement and the 
model prediction of the galaxy power spectrum $\Ph(k)$ and bispectrum 
$\BB(\kk_a,\kk_b,\kk_c)$, and $\CC(\pp)$ is the covariance matrix for the
measurements. For forecasting the parameter constraints in galaxy surveys
in section~\ref{sec:for} we have used the Fisher information matrix
(see, e.g., \cite{TETAHE97,WUROWE08})
\beeq
\label{aeq:fish}
F_{ij}=\left\langle{\partial^2\mathcal{L}\over\partial p_i
\partial p_j}
\right\rangle={1\over2}\up{Tr}\left[\CC^{-1}\CC_{,i}\CC^{-1}\CC_{,j}
+2\mm_{,i}^T\CC^{-1}\mm_{,j}\right]\simeq\sum_\kk{1\over\sigma_k^2}
{\partial\mu(\kk)\over\partial p_i}{\partial\mu(\kk)\over\partial 
p_j}~,
\eneq
where the first term involving the derivative of the
covariance matrix is smaller than the other, 
and hence we ignored the derivatives of the covariance 
matrix in computing the constraints.

When the assumed fiducial model is incorrect, our computation
of the mean and its covariance matrix will misrepresent the measurements,
providing biased best-fit parameters $\pp$ different from the true parameters
$\pp_t$. Assuming that our model is not far off from the true model, the
 best-fit parameters can be approximated as
$\pp=\pp_t+\bdv{\delta p}$, and the bias $\bdv{\delta p}$ 
in the parameter estimation
can be obtained from the relation that the best-fit parameters $\pp$ maximize
the likelihood in eq.~(\ref{eq:like}),
\bear
\label{eq:max}
0&=&\left\langle{\partial\mathcal{L}\over\partial p_i}(\pp)\right\rangle=
\up{Tr}\left[\CC^{-1}\CC_{,i}-\CC^{-1}\CC_{,i}\CC^{-1}\langle\dd\rangle
+\CC^{-1}\langle\dd_{,i}\rangle\right] \\
&=&\up{Tr}\left[\CC^{-1}\CC_{,i}\CC^{-1}(\CC-\CC_t)
-2\dmm^T\CC^{-1}\mm_{,i}
\right]+2\sum_j\delta p_j\up{Tr}\left[{1\over2}\CC^{-1}\CC_{,i}\CC^{-1}\CC_{,j}
+\mm_{,j}^T\CC^{-1}\mm_{,i}\right]~. \nonumber
\enar
The data matrix is $\dd=(\xx-\mm)(\xx-\mm)^T$, and the ensemble averages
of the data matrix and its derivative are
\bear
\langle\dd\rangle&=&\CC_t(\pp_t)+(\mm_t-\mm)(\mm_t-\mm)^T\simeq\CC_t
-\sum_j\CC_{,j}\delta p_j+\mathcal{O}(\delta p^2)~,\\
\langle\dd_{,i}\rangle&=&2\mm_{,i}\sum_j\mm_{,j}^T\delta p_j-
2\mm_{,i}\dmm^T~,
\enar
where $\CC_t$ is the true covariance matrix
and the mean is related to the true mean as
$\mm(\pp_t)=\mm(\pp)+\dmm$. Substituting these relations in eq.~(\ref{eq:max}),
we can obtain the bias in our parameter estimation as
\beeq
\delta p_j=\sum_i\left(\ff^{-1}\right)_{ij}\up{Tr}
\left[{1\over2}\CC^{-1}\CC_{,i}\CC^{-1}(\CC_t-\CC)+\dmm^T\CC^{-1}\mm_{,i}
\right]\simeq\sum_\kk{1\over\sigma_k^2}{\partial\mu(\kk)\over\partial p_i}
\delta\mu(\kk)\bigg/
\sum_\kk{1\over\sigma_k^2}
{\partial\mu(\kk)\over\partial p_i}{\partial\mu(\kk)\over\partial p_j}~,
\eneq
where we again ignored the derivative of the covariance matrix.

\end{document}